# Preliminary Tc calculations for iron-based superconductivity in NaFeAs, LiFeAs, FeSe and nanostructured FeSe/SrTiO$_3$ Superconductors

Wong Chi Ho[1, 2, 3*], Rolf Lortz[1, *]

[1]Department of Physics, Hong Kong University of Science and Technology, Hong Kong, China
[2]Department of Industrial and Systems Engineering, The Hong Kong Polytechnic University, Hong Kong, China
[3]Research Institute for Advanced Manufacturing, The Hong Kong Polytechnic University, Hong Kong, China

**Email address:**
chkhwong@ust.edu.hk (Wong Chi Ho), lortz@ust.hk (Rolf Lortz)
*Corresponding author

**Abstract:** Many theoretical models of iron-based superconductors have been proposed but Tc calculations based on the models are usually missing. We have chosen two models of iron-based superconductors in the literature and then compute the Tc values accordingly: Recently two models have been announced which suggest that superconducting electron concentration involved in the pairing mechanism of iron-based superconductors may have been underestimated, and that the antiferromagnetism and the induced xy potential may even have a dramatic amplification effect on electron-phonon coupling. We use bulk FeSe, LiFeAs and NaFeAs data to calculate the Tc based on these models and test if the combined model can predict the superconducting transition temperature ($T_c$) of the nanostructured FeSe monolayer well. To substantiate the recently announced xy potential in the literature, we create a two-channel model to separately superimpose the dynamics of the electron in the upper and lower tetrahedral plane. The results of our two-channel model support the literature data. While scientists are still searching for a universal DFT functional that can describe the pairing mechanism of all iron-based superconductors, we base on the ARPES data to propose an empirical combination of DFT functional for revising the electron-phonon scattering matrix in the superconducting state, which ensures that all electrons involved in iron-based superconductivity are included in the computation. Our computational model takes into account this amplifying effect of antiferromagnetism and the correction of the electron-phonon scattering matrix together with the abnormal soft out-of-plane lattice vibration of the layered structure, which allows us to calculate theoretical $T_c$ values of LiFeAs, NaFeAs and FeSe as a function of pressure that correspond reasonably well to the experimental values. More importantly, by taking into account the interfacial effect between an FeSe monolayer and its SrTiO$_3$ substrate as an additional gain factor, our calculated $T_c$ value is up to 91 K high, and provides evidence that the strong $T_c$ enhancement recently observed in such monolayers with $T_c$ reaching 100 K may be contributed from the electrons within the ARPES range.

## 1. Introduction

The pairing mechanism of the unconventional high-temperature superconductors (HTSC) remains one of the greatest unsolved mysteries of physics. All unconventional superconductors, including cuprates [1, 2] and iron-based HTSC [3, 4], but also heavy fermions [5] and organic superconductors [6], have in common that the superconducting phase occurs near a magnetic phase. Furthermore, their phase diagrams typically show at least one other form of electronic order, e.g. charge or orbital order [7, 8], a pseudogap phase [2], stripe order [2] or nematic order [9]. The proximity of the magnetic phases naturally suggests the involvement of magnetism [10]. In most theoretical approaches, spin fluctuations play a leading role [11, 12]. Alternative approaches consider e.g. excitonic superconductivity [13, 14], long-wavelength plasmonic charge fluctuations or orbital fluctuations [15-17].

It is generally assumed that the Cooper pairing in these superconductors cannot be described within a standard phonon-mediated scenario. However, this assumption is based only on the consideration of electron-phonon coupling on the Fermi surface only. The $T_c$ calculation based on the McMillan $T_c$ formula typically uses an approximation valid for classical low-Tc superconductors, where the superconducting electron concentration is only considered at the Fermi level. This approximation is no longer valid for high-temperature superconductors such as the iron-based superconductors, since high-energy phonons are excited at elevated temperatures, so that electron-phonon scattering influences the electron over a larger energy range around the Fermi energy. In the high temperature limit, where phonons are excited to the Debye energy, this energy interval becomes. Experimental ARPES data actually show that in iron-based superconductors electrons down to ~0.03-0.3eV below the Fermi energy are influenced by the onset of superconductivity [18-20]. In order to perform a comprehensive study of whether the electron-phonon coupling is related to the formation of Cooper pairs in iron-based superconductors or not, we decide to consider the true superconducting electron concentration in order to recalculate the electron-phonon coupling constant under antiferromagnetic background. Several studies offered an alternative scenario for iron-based superconductors, suggesting that the role of electron-phonon coupling had previously been underestimated against the antiferromagnetic (AF) background [21-23]. An explicit DFT calculation by B. Li et al [22] showed that the phonon softening of AFeAs (A: Li or Na) under AF background allows an increase of the electron-phonon coupling by a factor of ~2. While any orthogonal change of the phonon vector can be considered a phonon-softening phenomenon, the lattice dynamics studied by S. Deng et al [23] confirmed that out-of-plane lattice vibration amplifies electron-phonon scattering based on their first-principle linear response calculation. While the tetrahedral atom is better suited to attract electrons in terms of electronegativity, the vertical displacement of the lattice Fe transfers the charge of the electron to the tetrahedral regions to generate an additional xy potential [21]. S. Coh et al [21] calibrated the GGA+A functional, which made it possible to bring the simulation results much closer to the experiments [21,24]. The calibrated ab-initio method explicitly demonstrates the occurrence of the induced xy potential from the out-of-plane lattice dynamics in the AF background that increase the electron-phonon scattering matrix by this factor of ~2 (abbreviated as ratio $R_{ph}$). More importantly, they provide an analytical model [21] to explain why the electron-phonon scattering computed by the ab-initio method is always increased by a ratio of ~2 under the effect of the spin density wave (abbreviated as ratio $R_{SDW}$).

The pairing strength of iron-based superconductivity can be enhanced significantly with the help of nanostructuring [25-28]. The layer structure of FeSe makes it possible to grow monolayers of FeSe epitaxially on a substrate. In 2013, superconductivity was reported with a record $T_c$ of 70 K on monolayer FeSe on a $SrTiO_3$ substrate [25], which was later increased to 100 K [26]. Despite the complexity of the electronic phase diagram of iron–based superconductors, which suggests the presence of additional broken symmetries besides the broken U(1) gauge symmetry of the superconducting state and thus an unconventional pairing mechanism, recent works have suggested that the role of electron-phonon coupling could play a certain role in the superconducting mechanism of iron-based superconductors [27-29], although there is clear evidence that magnetic fluctuations must be taken into account. The high transition temperature of the monolayer FeSe on a $SrTiO_3$ substrate gives further indications of the importance of electron-phonon coupling. While growing FeSe films on graphene substrate suppresses $T_c$ [30], the giant enhancement of $T_c$ is likely activated by the $SrTiO_3$ substrate, where the interfacial contribution cannot be ignored. Strong electron–phonon coupling at the interface of FeSe/$SrTiO_3$ has been identified in ARPES data [19], with electrons located 0.1-0.3eV below the Fermi level involved in superconductivity. Although the FeSe phonons do not depend on the thickness of the FeSe material, the unusual phonon [31,32] such as F-K phonon across the interface may be responsible for the high $T_c$ [32]. According to the experiment by S. Zhang *et al* [32], the F-K phonons of the FeSe/$SrTiO_3$ surface show new energy loss modes and the line width is widened compared to bare $SrTiO_3$. In this article, we revise the superconducting electron concentration and use an ab-initio approach to examine if the Tc values of LiFeAs, NaFeAs and FeSe as a function of pressure can be calculated reasonably by taking into account the $R_{ph}$ and $R_{SDW}$ factors...etc. If succeed, we use this model to test whether such an approach can be applied to the ~100K superconductivity in the nanostructured FeSe/$SrTiO_3$. *Not all mechanisms of iron-based superconductivity have been encountered in this work because the unified theory of iron-based superconductors remains an open question. We only apply mathematical techniques to convert the two models from the literature into Tc values which may be important to find out the possible mechanism of iron-based superconductors.*

**2. Computational Methods**

As a starting point, the electronic properties) of all compounds investigated in this article are computed by the spin-unrestricted GGA-PBE functional (unless otherwise specified) [32-36] in Wien2K. The SCF tolerance is 1x10$^{-5}$eV and the interval of k-space is 0.025(1/Å). The maximum SCF cycle is 1000. The magnetism and phonon data are calculated by CASTEP. Finite displacement mode is chosen where the supercell defined by cutoff radius is 5Å and the interval of the dispersion is 0.04(1/Å). Ultrasoft pseudopotential is assigned and density mixing is chosen to be the electronic minimizer [32-36]. The experimental lattice parameters are used [37, 38]. In this article, only Fe and As atoms are imported for the 111-type compounds.

Instead of calibrating 'A' in the GGA+A functional, which entails an enormous computational cost and time-consuming experimental effort [21, 39, 40], we propose a two-channel model to more easily model the induced $xy$ potential, where the upper tetrahedral plane is called channel 1 and the lower tetrahedral plane is called channel 2, respectively. We apply the superposition principle to separately calculate the induced $xy$ potentials induced by channel 1 and 2. Our two-channel model has fulfilled an assumption that the probability of finding an Fe atom moving in the $+z$ and $-z$ directions is equal, but that their vibrational amplitudes never cancel each other out. This assumption is justified by Coh et al whose explicit calculation confirms that the iron-based system consists of an out-of-phase vertical displacement of iron atoms, with first adjacent iron atoms moving in opposite directions [21].

We define $R_{ph} = \dfrac{0.5\left(DOS_1^{XY} + DOS_2^{XY}\right)}{DOS_{12}^{XY}}$ where $DOS_c^{XY}$ is the average electronic density of states within the ARPES range. The index $c$ refers to the channel index.

Define $F(\omega)$ is the phonon density of state as a function of frequency $\omega$ and the integral $\int d^2 p_F$ is taken over the Fermi surface with the Fermi velocity $v_F$. The Eliashberg function is written as [41]

$$\alpha^2 F(\omega) = \int \dfrac{d^2 p_F}{v_F} \int \dfrac{d^2 p_F{'}}{(2\pi\hbar)^3 v_F{'}} \sum_v g_{pp'v}^2 \delta\left(\omega - \omega_{p-p'v}\right) / \int \dfrac{d^2 p_F}{v_F}$$

The electron-phonon matrix elements is given by $g_{pp'v} = \sqrt{\dfrac{\hbar}{C\omega_{p-p'v}}} g_v(p,p')$ where $\int \psi_p^* u_i \cdot \nabla V_{XY} \psi_p dr$ is abbreviated as $g_v(p,p')$ and C is the material constant related to lattice [41]. The $u_i$ and $V_{XY}$ represent the displacement of ion relative to its equilibrium position and the ionic potential. The $\psi_p^* \psi_p$ is the electronic probability density in the non-magnetic state. The resultant ionic interaction $V_{ion}^{XY}$ on the $XY$ plane due to the abnormal phonon is calculated by multiplying the ionic potential by $R_{ph}$, i.e. $V_{ion}^{XY} = V_{XY} \cdot R_{ph}$. Moreover, the antiferromagnetic interaction along the $XY$ plane amends the electronic probability density which fulfills $\phi_p^* \phi_p \sim \psi_p^* R_{SDW} \psi_p$ where the spin density wave factor $R_{SDW}$ is directly obtained from the ab-initio calculation. Rearranging the mathematical terms yields the electron-phonon matrix element. as

$$g_{pp'v} = \sqrt{\dfrac{\hbar}{C\omega_{p-p'v}}} \int u_i \cdot \nabla\left(V_{XY} R_{ph}\right) \psi_p^* R_{SDW} \psi_p dr = \sqrt{\dfrac{\hbar}{C\omega_{p-p'v}}} \int \phi_p^* u_i \cdot \nabla V_{ion}^{XY} \phi_p dr$$

To derive a superconducting transition temperature from the computed parameters, we use the McMillan T$_c$ formula [41]. Due to the high transition temperatures, the electron-phonon scattering matrix takes into account the full electronic DOS in range of $E_F - E_{Debye}$ to $E_F$ and not only the value at Fermi level (i.e. effective electronic DOS is increased). Here we consider the fact that $E_{Debye}$ represents the upper limit of the phonon energies that can be transferred to electrons, and at the high transition temperatures of Fe-based superconductors, contributions from high energy phonons become important in the electron-phonon scattering mechanism, as opposed to classical low-Tc superconductors. Although this approach is a simple consequence of the conservation of energy, it is supported by experiments: A shift of the spectral weight between the normal and the superconducting state is clearly visible in the photoemission spectra below the superconducting energy gap of various iron-based compounds in an energy range of ~30 - 60 meV below the Fermi energy [18-20]. This energy range is approximately in the order of the Debye energy.

In BCS superconductors, the electrons on the Fermi surface condense into the Bose-Einstein superconducting state where the total number of electrons on the Fermi surface equals to the total number of electrons on the superconducting state. Hence, the theoretical Tc of BCS superconductors remain the same if we substitute either the electronic DOS on the Fermi level or the electronic DOS of the condensed Bose-Einstein state. However, the situation is different in iron-based superconductivity where the electrons located between $E_F - E_{Debye}$ and $E_F$ transfer energy to the electrons in the Bose-Einstein superconducting states. When this happens, we have to revise the resultant electron-phonon scattering matrix in the condensed Bose-Einstein state. Bose-Einstein statistic favors more electrons occupying in the superconducting state. The electrons within ARPES range increases the effective electronic DOS in the condensed Bose-Einstein state indirectly. Those electrons within the ARPES range cannot be excited to the Fermi surface due to electrostatic repulsion. However, those electrons have another route to follow the Bose-Einstein distribution which can be argued as a reason why those electrons disappear below the Fermi level.

The computation of band structure produces discrete (E,k) points where E and k are the energy and the wavevector of electron.

The ratio of the electron-phonon scattering matrix is $R_g = \dfrac{\sum_{-\infty}^{E_F} g(E)\delta_A(E) / \sum_{-\infty}^{E_F} \delta_{counter}(E)}{\sum_{-\infty}^{E_F} g(E)\delta_B(E) / \sum_{-\infty}^{E_F} \delta_{counter}(E)}$ which is called the

ARPES factor. The $\delta_A(E)$ is 1 if $(E_F - E_D) \leq E \leq E_F$. Similarly, the $\delta_B(E) = 1$ if $E = E_F$. Otherwise $\delta_A(E) = \delta_B(E) = 0$. $\sum_{-\infty}^{E_F}\delta_{counter}(E)$ gives the total number of (E,k) points in the range $-\infty \leq E \leq E_F$. $\sum_{-\infty}^{E_F}\delta_A(E)/\sum_{-\infty}^{E_F}\delta_{counter}(E)$ or $\sum_{-\infty}^{E_F}\delta_B(E)/\sum_{-\infty}^{E_F}\delta_{counter}(E)$ is the percentage of electrons contributed to the $R_g$ factor. To make a fair comparison, the interval of k space in the numerator and denominator of $R_g$ are essentially the same. The $R_g$ factor controls the proportion of electrons scattered below the Fermi level.

Due to the fact that the superconducting transition temperatures are low, we calculate the mean occupation number $f(E)$ in the Fermi-Dirac statistic at low temperatures (T < 100K), where $f(E_F)$ and $f(E_F - E_{Debye})$ are 0.5 and ~0.5005, respectively. If $DOS(E_F)/DOS(E_F - E_{Debye}) \sim 1$, $f(E_F)/f(E_F - E_{Debye}) \sim 1$ and $E_F \gg E_{Debye}$, the tiny offset in the mean occupation number may allow the Eliashberg function to approximately obey the following form.

$$\alpha_{PS}^2 F(\omega) \sim \left\langle \sum_{V_F - V_{Debye}}^{V_F} \int \dfrac{d^2 p_E}{v_E} \right\rangle \left\langle \sum_{V_F - V_{Debye}}^{V_F} \int \dfrac{d^2 p_E'}{(2\pi\hbar)^3 v_E'} \right\rangle \sum_{v} \delta(\omega - \omega_{p-p'v}) \left| \sqrt{\dfrac{\hbar}{C\omega_{p-p'v}}} \int u_i \cdot \nabla(V_{XY}R_{ph})\psi_p^* R_{SDW} R_g \psi_{p'} dr \right|^2 / \left\langle \sum_{V_F - V_{Debye}}^{V_F} \int \dfrac{d^2 p_E}{v_E} \right\rangle$$

where $v_E \in (v_F - v_{Debye}, v_F)$ and the velocity $v_{Debye}$ is converted from the Debye energy. $\sum_{V_F - V_{Debye}}^{V_F} \int \frac{d^2 p_E}{v_E}$ is the sum of the surface integral $\int \frac{d^2 p_E}{v_E}$ at different electron energies within the ARPES range. The form of the antiferromagnetically amplified electron-phonon coupling is expressed as $\lambda_{PS}^{Coh} \sim 2\int \alpha_{PS}^2 \frac{F(\omega)}{\omega} d\omega$ where $\alpha_{PS}^2 \sim \alpha_{E_F}^2 R_{Ph}^2 R_{SDW}^2 R_g^2$. The $\alpha_{E_F}$ is the average square of the electron phonon scattering matrix on the Fermi surface [41]. In the case of strong coupling, the renormalized electron-phonon coupling is expressed as $^*\lambda_{PS}^{Coh} = \frac{\lambda_{PS}^{Coh}}{\lambda_{PS}^{Coh} + 1}$ [42].

When the pairing strength is calculated by the spin-unrestricted GGA-PBE functional without using the AF Ising Hamiltonian, this approach is defined as 'traditional combination of DFT functional'. On the other hand, we propose an 'empirical combination of DFT functional'. i.e. the average electron-phonon coupling in multi-energy layers is computed by the spin-restricted GGA-PBE functional [34] and further corrected by the AF Ising Hamiltonian. To include the magnetic effect, this AF Ising Hamiltonian must be acquired by the spin-unrestricted GGA-PW91 functional.

The pairing strength formulas of LiFeAs (111-type), NaFeAs (111-type) and FeSe (11-type) under pressure are given as $\lambda_{11}^{111} = {}^*\lambda_{PS}^{Coh} f_{11}^{111}(E_{ex})$ where $f_{11}^{111}(E_{ex}) \sim \frac{[M_{Fe} M_{Fe} E_{co}]_{P>0}}{[M_{Fe} M_{Fe} E_{co}]_{P=0}}$. The ratio $f_{11}^{111}(E_{ex})$ monitors the pressure dependence of the AF energy at each external pressure $P$ where $E_{co}$ is the exchange-correlation coupling. We use $f_{11}^{111}(E_{ex})$ to correct the antiferromagnetism under pressure instead of recalculating the $R_{SDW}^2$. The Debye temperature of the FeSe/SrTiO$_3$ is replaced by the vibrational energy of F-K phonon across the interface [32]. The pairing strength is substituted into the McMillian $T_c$ formula [27], which includes the enhanced electron-phonon scattering matrix elements:

$$T_c = \frac{T_{Debye}}{1.45} \exp\left( \frac{-1.04(1 + \lambda_{11}^{111})}{\lambda_{11}^{111} - \mu^*(1 + 0.62\lambda_{11}^{111})} \right)$$

## 3. Results

The atomic spring constants between the FeFe bond $k_{FeFe}$ and FeSe bond $k_{FeSe}$ in the iron-based superconductors are compared. Our DFT calculation shows that $k_{FeSe}/k_{FeFe} \sim 0.25$, while the $k_{FeAs}$ is almost 2 times stronger than $k_{FeSe}$. As the atomic spring constants of the tetrahedral bonds are comparable to the FeFe bond, appearing the orthogonal phonon is feasible. Our two-channel model demonstrates that the induced $xy$ potential is good enough be emerged at 'GGA-PBE' level. We calculated that the electron-phonon scattering matrix of FeSe under the induced $xy$-potential is amplified by $R_{ph}=2.8$. While the accuracy of our two-channel model is comparable to the $R_{ph}=2.2$ obtained from the calibrated GGA+A functional [21], we determine $R_{ph}$ of NaFeAs and LiFeAs to be 1.97 and 1.8, respectively. The pressure dependence on $R_{ph}$ is less than ~5% due to $c \gg a$.

A critical parameter in any ab-initio approach is the value of the renormalized Coulomb pseudopotential. Figure 1 estimates the error of the theoretical $T_c$ by tuning $\mu^*$. Despite the calculation of $\mu^*$ as a function of Debye temperature and Fermi level [42] may not be very accurate in such a strongly correlated electron system [43], it has been argued that for the most Fe-based superconductors $\mu^*$ should be 0.15-0.2 [44]. The error of our $T_c$ calculation due to the uncertainty of $\mu^*$ is within ~15%. In this letter we choose the value ($\mu^*$=0.15) of the Coulomb pseudopotential to calculate the $T_c$ of LiFeAs, NaFeAs and FeSe to make a fair comparison.

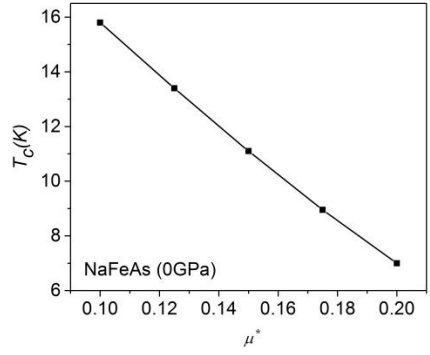

**Figure 1.** *The theoretical $T_c$ of NaFeAs varies slightly with the Coulomb pseudopotential. Our calculated $\mu^*$-value of the uncompressed NaFeAs is 0.13.*

Figure 2a shows that our approach can generate the theoretical $T_c$ values in an appropriate range. The ARPES data confirms that LiFeAs and FeSe require the use of the $R_g$ factor, while the NaFeAs does not [18, 20, 45]. The theoretical $T_c$ of NaFeAs at 0GPa and 2GPa are 11K and 12.5K, respectively [46]. The antiferromagnetically enhanced electron-phonon interaction on the Fermi surface and the AF exchange Hamiltonian compete in the compressed NaFeAs as illustrated in Figure 2b. We observe that the antiferromagnetism is slightly weaker at finite pressure, but the antiferromagnetically assisted electron-phonon coupling on the Fermi layer is increased almost linearly al low pressure. We show the steps to estimate the $T_c$ of NaFeAs at 0GPa as an example. After activating the spin-unrestricted mode, the $R_{SDW}^2$ is 1.625. The antiferromagnetically assisted electron-phonon coupling on the Fermi surface is $\lambda_{PS}^{Coh} = \lambda_{E_F} R_{SDW}^2 R_{ph}^2 R_g^2 = (0.13)(1.625)(1.97^2)(1^2) = 0.819$ and $\mu^* = 0.15$.

According to the McMillian $T_c$ Formula, the $T_c$ becomes

$$T_c = \frac{T_{Debye}}{1.45} \exp\left(\frac{-1.04(1+\lambda_{11}^{111})}{\lambda_{11}^{111} - \mu^*(1+0.62\lambda_{11}^{111})}\right) = \frac{385}{1.45}\exp(-3.19) = 10.9K$$

We compare our theoretical $T_c$ by substituting the raw data of other groups [15, 21], their calculated $\lambda_{E_F}^{AF}$ is 0.39 [15] and the induced *xy* potential by the out-of-plane phonon reinforces the electron-phonon coupling matrix by 2.2 [21].

$$\lambda_{PS}^{Coh} = \lambda_{E_F}^{AF} R_{ph}^2 R_g^2 = (0.39)(2.2^2)(1^2) = 1.88$$

After renormalization, these two couplings are softened to

$$\lambda_{11}^{111} = {}^*\lambda_{PS}^{Coh} = 1.88/(1.88+1) = 0.652$$

And the renormalized Coulomb pseudopotential $\mu_{re}^* = \dfrac{\mu^*}{1+\lambda_{PS}^{Coh}} = 0.15/(1.88+1) = 0.052$.

The theoretical $T_c$ becomes

$$T_c = \frac{T_{Debye}}{1.45} \exp\left(\frac{-1.04(1+\lambda_{11}^{111})}{\lambda_{11}^{111} - \mu_{re}^*(1+0.62\lambda_{11}^{111})}\right) = \frac{385}{1.45}\exp(-2.97) = 13.6K$$

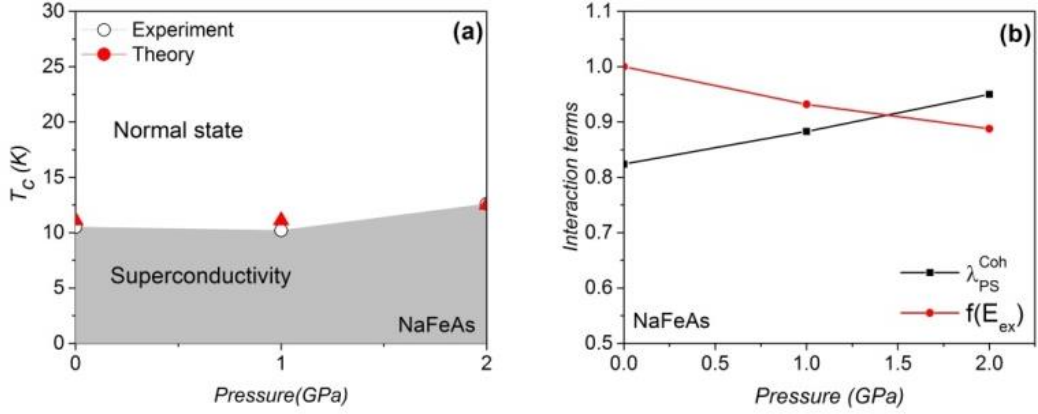

***Figure 2.*** *a The theoretical and experimental [46] $T_c$ values of NaFeAs. b The antiferromagnetically assisted electron-phonon coupling on the Fermi surface and the AF energy as a function of pressure.*

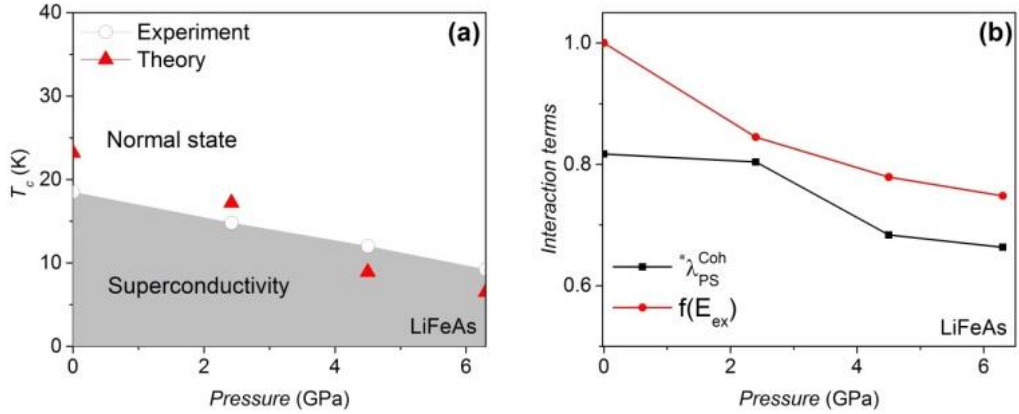

***Figure 3.*** *a The theoretical and experimental [47] $T_c$ values of LiFeAs are consistent. b The antiferromagnetically assisted electron-phonon coupling and the AF exchange Hamilton under pressure. The $R_{SDW}^2$ equals to 1.75.*

Our calculated value of the electron-phonon coupling on the Fermi surface of the uncompressed LiFeAs is ~0.1 [48] but the magnetic amplification factors increase the pairing strength to 0.82, remarkably. The Debye temperature $T_{Debye}$ of LiFeAs remains at ~385K below 8GPa [49], as shown in Table 1. A reduction of the theoretical $T_c$ is also observed in the compressed LiFeAs and the weakening effect of $^*\lambda_{PS}^{Coh}$ and $f_{11}^{111}(E_{ex})$ under pressure are identified, as shown in Figure 3b. In compressed FeSe [24], however, a gain in $f_{11}^{111}(E_{ex})$ is observed that triggers the increase of $T_c$ under pressure (Figure 4). It should be noted that our approach is a mean field approach and we treat the spin fluctuations as being proportional to the mean field Hamiltonian. The vanishing of the macroscopic AF order observed in real samples is due to the strong fluctuation effects in these layered compounds. The magnetism considered here in the non-magnetic regimes of the phase diagrams is of a fluctuating microscopic nature. The optimized pairing strength of LiFeAs and FeSe is achieved at a pressure of 0GPa and 0.7GPa, respectively. The differences between DOS($E_F$–$E_{Debye}$) and DOS($E_F$) in LiFeAs and FeSe are less than 4%. The $R_g$ factor in LiFeAs is reduced with pressure, but the $R_g$ factor of FeSe is optimized at medium pressure (see Table 1-2).

***Table 1.*** *The DFT parameter of LiFeAs. The $R_g$ factor is compiled by the 'empirical combination of DFT functional'.*

| **P/GPa** | ***a*** **(Å)** | ***c*** **(Å)** | **FeAs length (Å)** | **$R_g$** | **$T_{Debye}$ (K)** |
|---|---|---|---|---|---|
| 0 | 3.769 | 6.306 | 2.44 | 2.66 | 385.00 |
| 2.4 | 3.745 | 6.134 | 2.42 | 2.38 | 385.25 |
| 4.5 | 3.723 | 5.985 | 2.35 | 1.67 | 385.5 |
| 6.3 | 3.702 | 5.918 | 2.33 | 1.56 | 385.75 |

**Table 2.** *The DFT parameter of FeSe. The $R_g$ factor is simulated by the 'empirical combination of DFT functional'.*

| P/GPa | $a$ (Å) | $c$ (Å) | FeSe length(Å) | $R_g$ | $T_{Debye}$ (K) |
|---|---|---|---|---|---|
| 0 | 3.767 | 5.485 | 2.390 | 3.04 | 240 |
| 0.7 | 3.746 | 5.269 | 2.388 | 2.05 | 256 |
| 2.0 | 3.715 | 5.171 | 2.384 | 4.92 | 274 |
| 3.1 | 3.698 | 5.114 | 2.382 | 2.50 | 290 |

**Table 3.** *The DFT parameter of NaFeAs. The $R_g$ factor is computed by the 'empirical combination of DFT functional'.*

| P/GPa | $a$ (Å) | $c$ (Å) | FeAs length(Å) | $R_g$ | $T_{Debye}$(K) |
|---|---|---|---|---|---|
| 0 | 3.929 | 6.890 | 2.400 | 1.00 | 385.0 |
| 1 | 3.914 | 6.833 | 2.388 | 1.00 | 385.5 |
| 2.0 | 3.900 | 6.777 | 2.376 | 1.00 | 386.0 |

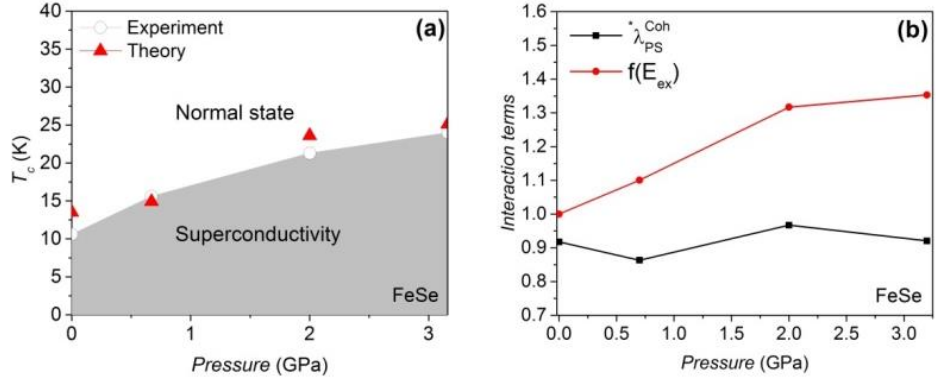

**Figure 4.** *a Both theoretical and experimental [24] $T_c$ values increase with pressure. b The pressure dependence of the antiferromagnetically assisted electron-phonon coupling and the AF interaction. The $R_{SDW}^2$ at 0GPa is 1.59.*

Based on the successful Tc calculation of the bulk FeSe, LiFeAs and NaFeAs, we start our journey to acquire the theoretical $T_c$ of monolayer FeSe on a SrTiO$_3$ substrate step by step using the model of an antiferromagnetically-enhanced electron-phonon coupling. The flowchart is shown in Figure 5. After geometric optimization, the angles of the unit cell are 89.81°, 90.88°, 89.05°, with a tiny internal shear force being captured. The relaxed tetrahedral angle of Fe-Se-Fe is 108 degrees. The antiferromagnetic energy of FeSe can be amplified by low dimensionality when it is deposited in form of a monolayer on SrTiO$_3$ [26]. Compared to an FeSe monolayer without substrate, the FeSe film on SrTiO$_3$ shows an increased exchange correlation energy of ~16%. Apart from this, the local Fe moment in the isolated FeSe film is only ~0.5$\mu_B$. However, the contact to SrTiO$_3$ amplifies the local Fe moment up to ~1.3$\mu_B$. Our calculated the electron-phonon coupling on the Fermi surface without any amplification factor is $\lambda_{Fermi} = 0.12$. Based on our simulation, the antiferromagnetism of FeSe/SrTiO$_3$ is still as strong as of the FeSe monolayer without substrate. Hence the simultaneous occurrence of antiferromagnetism and tetrahedral atoms makes the Coh factor unavoidable. The analytical result of $C_{AF}$ = 2 is used and our calculated $C_{Ph}$ in FeSe/SrTiO$_3$ is 2.9. After amplification of the Coh factor, the theoretical $T_c$ is only 14K. However, a massive enhancement of the pairing strength can be observed when the interfacial F-K phonon is involved [32]. The F-K phonon actuated via the interface contributes the vibrational energy of ~100meV (~1159K) [32]. With this enormous Debye temperature, the theoretical $T_c$ is increased to 69K, although the electron-phonon interaction is limited to the Fermi energy. In ARPES data it is evident that a shift of spectral weight occurs in the superconducting state 0.1~0.3eV below the Fermi level [19], which means that electrons in this energy range are affected by electron-phonon scattering as a result of the high phonon frequencies. This means that electrons in this energy range contribute to superconductivity, since the high phonon frequencies can scatter them up to the Fermi energy and need to be considered in the McMillan formula, and not only those at the Fermi energy as in the usual approximation applied to classical low-$T_c$ superconductors. The superconducting electron concentration is thus corrected and the average electron-phonon scattering matrix in these multi-energy layers is 1.96 times higher than the matrix considering only the Fermi level.

This is the last factor with which our theoretical $T_c$ can reach 91K, which corresponds quite well to experimental $T_c$ of 100 K.

The pairing strength is renormalized as

$$^*\lambda_{PS} = \frac{\lambda_{PS}}{\lambda_{PS}+1} = \frac{R_g^2 C_F^2 \lambda_{Fermi}}{R_g^2 C_F^2 \lambda_{Fermi}+1} = \frac{(1.96^2)(2^2)(2.99^2)(0.12)}{(1.96^2)(2^2)(2.99^2)(0.12)+1} = 0.942$$

The pseudopotential is diluted as $\mu^* = \dfrac{\mu}{1+\lambda_{PS}} = \dfrac{0.15}{1+(1.96^2)(2^2)(2.99^2)(0.12)} = 0.0085$

We substitute all parameters into the McMillian $T_c$ formula,

$$T_c = \frac{T_{Debye}}{1.45}\exp\left(\frac{-1.04(1+{}^*\lambda_{PS})}{{}^*\lambda_{PS}-\mu^*(1+0.62{}^*\lambda_{PS})}\right) = \frac{1159}{1.45}\exp\left(\frac{-1.04(1+0.942)}{0.942-0.0085(1+0.62(0.942))}\right) = 91K$$

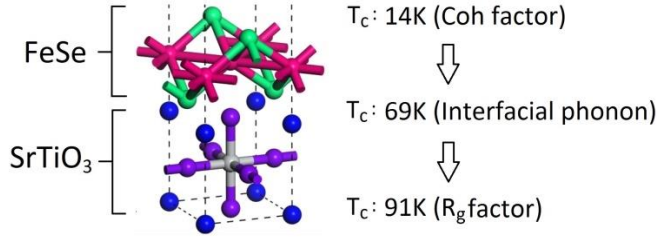

Figure 5: The local region of the unit cell. A FeSe monolayered film is deposited on a SrTiO$_3$ monolayer to form a composite. The vacuum distance D above the composite is ~52Å. Our theoretical $T_c$ values are shown after the amplifications of interfacial F-K phonon, Coh factor and the ARPES factor (or $R_g$ factor) [21,26,32].

### 4. Discussion

<u>4.1. Are ARPES data and the Coh factors important to IBSC?</u>

The pure FeAs layer in the 111-type, 1111-type and 122-type Fe-based superconductors is believed to trigger superconductivity [50-51]. The investigation of the pure FeAs layer without the Li and Na atoms in the simulation can show the bare pairing strength. The $T_c$ vs. pressure of the NaFeAs is not as sensitive as for the other materials. The reason for this is that the increase of $^*\lambda_{PS}^{Coh}$ and the decrease of $f_{11}^{111}(E_{ex})$ almost cancel out the variation in the pairing strength. The unusually high $T_c$ in the LiFeAs and FeSe at 0GPa is mainly due to the $R_{ph}$, $R_{SDW}$ and $R_g$ factors. Our approach confirms that the reduction of $T_c$ in compressed LiFeAs is mainly due to the decreases in $^*\lambda_{PS}^{Coh}$ and AF energy as a function of pressure. Conversely, the magnetic moment of Fe in FeSe increases under compression, resulting in an increase in AF energy under pressure. As a result, the increase of $T_c$ in compressed FeSe is observed. The $R_g$ factor is minimized at high pressure since the kinematics of electrons below the Fermi level is more restricted under pressure. Our simulation shows that the variation of the induced $xy$ potential is less than ~3% for the electrons at ~100meV below the Femi level and therefore the use of the $R_g$ factor in LiFeAs and FeSe is justified.

We correct the pairing strength at high pressures with help of the AF Ising Hamitonian. In the following we compare the $T_c$ when the $R_{SDW}$, $R_g$ and $R_{ph}$ are calculated by the spin-unrestricted GGA-PBE functional at high pressures or simply called the 'traditional combination of DFT functional'. Despite the 'traditional combination of DFT functional' provides an accurate theoretical $T_c$ at ambient pressure, the error of $T_c$ is significant at high pressures. We demonstrate this for the case of FeSe in Table 4. In this approach we do not use the AF Ising Hamiltonian at finite pressure because magnetism is already considered. Since 2008, the $R_g$ factor was missing in the calculation of the electron-phonon coupling constant. However, Table 5 confirms that the consideration of the electron-phonon coupling on the Fermi surface is not sufficient to argue whether iron-based superconductivity is mediated by phonons. If the $R_g$ factor really participates in iron-based superconductivity, the abnormal distribution of electrons below the Fermi level should be given a larger range when the $T_c$ of the iron-based superconductor is higher. This argument is supported by the ARPES data of the 100K 2D FeSe/SrTiO$_3$ [19] with the parameters shown in Table 6. For these ~10K-30K iron-based superconductors, the electrons located at 0.03eV-0.06eV below the Fermi level are affected by superconductivity [28, 30]. However, the electrons in the 100K 2D FeSe/SrTiO$_3$, which are located in a much wider range of 0.1eV-0.3eV below the Fermi level, participate superconductivity [19]. The theoretical $T_c$ of the 2D FeSe/SrTiO$_3$ reaches 91K only if the $R_g$ factor is considered.

An empirical rule is that the $T_c$ of the iron-based superconductor is optimized when the tetrahedral angle is close to 109.5 degree [52]. When the FeSe monolayer is attached to the SrTiO$_3$, the tetrahedral angle is changed from 103 degrees to 108 degrees and the $T_c$ is benefits. However, all these antiferromagnetic and tetrahedral effects cannot explain the high $T_c$ near 100K until the interface properties are considered [32]. Despite the Debye temperature of the FeSe phonons (~250K) shows no significant size effect, an energetic F-K phonon carrying energy of 100meV (~1159K) was observed at the interface between the FeSe film and SrTiO$_3$ [32]. Since the 3D and 2D FeSe phonon are almost identical [32], the out-of-plane phonon from the tetrahedral sites should amplify the electron-phonon coupling of FeSe/SrTiO$_3$ by the same factor 2. Assuming that the F-K phonon and FeSe phonon interact with electrons simultaneously, two Debye energies, i.e. from the FeSe phonons and the F-K phonons, may influence the Cooper pairs. The two-fluid model, however, ensures that the onset $T_c$ is always related to the mechanism that gives the strongest pairing strength and therefore choosing 1159K as the Debye temperature is justified.

The ARPES data of FeSe/SrTiO$_3$ show that the electrons in a wide range below the Fermi level ($\Delta E$ ~0.1 - 0.3eV) participate in superconductivity [19]. A question may be asked: Which energy source causes this shift of spectral weight? The F-K phonon may be one of the options since the $E_{Debye}$ is ~0.1eV [32]. Would it be exchange coupling? The exchange-correlation energy $E_{co}$ of FeSe/SrTiO$_3$ is also ~0.1-0.2eV. However, we believe that the F-K phonon is the energy source to generate this shift of spectral weight in FeSe/SrTiO$_3$. To support our argument, we revisit the ARPES results [18,20], where the bulk iron-based superconductors carrying $E_{co}$ ~ 0.1eV display a shift of spectral weight at $\Delta E$ ~ 30 - 60 meV below the Fermi level. If the shift is caused by the exchange-correlation energy, $\Delta E$ and $E_{co}$ should be comparable in the bulk iron-based superconductors, but this is not the case. If the exchange correlation energy is not the correct answer, we re-investigate the magnitude of $E_{Debye}$. Interestingly, the narrower range $\Delta E$ ~ 30 - 60 meV is comparable to the Debye temperature [53,54] of bulk iron-based superconductors. With this, we believe that $\Delta E$ ~ $E_{Debye}$ is unlikely to be a coincidence. The shift of spectral weight in ARPES in iron-based superconductors is thus likely triggered by phonon-mediated processes. After revising the electron concentration in the superconducting state, our calculated $T_c$ is further increased to 91K. We have verified that the Coh factor is only reduced by ~3% at $E_F$ - 100meV.

### 4.2. Would the errors in Tc be rescued by the effects of nematicity and spin-orbital coupling?

On the Fermi surface, a nematic order may be observed in various iron-based superconductors [52,55] and the electron-electron interaction should be influenced accordingly. Although our approach does not consider the nematic order, our approach averages the electron-phonon coupling between $E_F$ - $E_{Debye}$ and $E_F$, which pale the contribution from to the nematic order at the Fermi surface. From a mathematical point of view, the $\alpha_{PS}$ is calculated by $\alpha_{E_F} C_F R_g$, where the Coh factor $C_F$ is a constant. The $\alpha_{E_F}$ is directly proportional to the $|g(E_F)|$. If the nematic order changes the $g(E_F)$ value, the $R_g$ factor cancels the nematic contribution because the $R_g$ is inversely proportional to $|g(E_F)|$. The numerator of $R_g$ contains the average electron-phonon scattering matrix in multi-energy layers, where the Fermi energy is only one of them. Under these circumstances, the error of $\alpha_{PS}$ from neglecting the nematic effect is relatively small (The variation of Tc enhanced by nematic phase in S-doped FeSe is just a few kelvin! If the nematic phase is encountered in our approach, this may help increasing the calculated T$_c$ to 100K, but the T$_c$ calculation based on the concept of nematic phase is still an open question) and our $T_c$ calculation should remain accurate. The spin-orbital coupling SO may be a reason to cause the unusually high Tc in FeSe/SrTiO$_3$ due to the heavy elements in SrTiO$_3$. If the effect of SO is taken into account, the calculated T$_c$ may move even closer to the experimental value. Additionally, another source of error in the T$_c$ of FeSe/SrTiO$_3$ may be caused by the thickness of SrTiO$_3$ used in the simulation. The theoretical T$_c$ of FeSe film may increase after the thickness of SrTiO$_3$ increases.

***Table 4.*** *The theoretical $T_c$ of FeSe at different pressures. Theoretical $T_c$ (A) is obtained from the traditional combination of DFT functional. Theoretical $T_c$ (B) is estimated from the empirical combination of DFT functional.*

| FeSe | Experimental $T_c$ | Theoretical $T_c$(A) | Theoretical $T_c$(B) |
|---|---|---|---|
| 0GPa | 11K | 13K | 12K |
| 0.7GPa | 16K | 4K | 15K |
| 2GPa | 20K | 3K | 22K |

***Table 5.*** *Effect of $R_g$ factor on theoretical $T_c$ values. The 'empirical combination of DFT functional' is used.*

| FeSe | Experimental $T_c$ | Theoretical $T_c$ (Without $R_g$ factor) | Theoretical $T_c$ (With $R_g$ factor) |
|---|---|---|---|
| 0GPa | 11K | 3K | 12K |
| 0.7GPa | 16K | 6K | 15K |
| 2GPa | 20K | 8K | 22K |
| LiFeAs | Experimental $T_c$ | Theoretical $T_c$ (Without $R_g$ factor) | Theoretical $T_c$ (With $R_g$ factor) |
| 0GPa | 19K | 2K | 23K |
| 2.4GPa | 15K | 7K | 17K |
| 4.5GPa | 13K | 8K | 9K |
| 6.3GPa | 10K | 4K | 7K |

***Table 6.*** *The simulation parameters of FeSe/SrTiO$_3$. The unit cell of FeSe/SrTiO$_3$ occupied the volume of 3.8197 Å x 3.8698 Å x 5.9540 Å. The layer-to-layer distance D is 52.484Å*

| a (Å) | b (Å) | c (Å) | D (Å) | $\lambda_{Fermi}$ | $R_g$ | Debye(K) |
|---|---|---|---|---|---|---|
| 3.8197 | 3.8698 | 5.9540 | 52.484Å | 1.6 | 1.96 | 1159 |

4.3. The universal theory of IBSC remains an open question

The $T_c$ acquired by the 'traditional combination of DFT functional' fails at high pressures mainly because $R_g$ is excessively suppressed. To monitor electron-phonon coupling under pressure, the use of the 'empirical combination of DFT functional' is a better choice. Although the accuracy of GGA-PBE functional may not be perfect, we empirically correct the numerical output value $\lambda_{11}^{111}$ directly via the AF Ising Hamiltonian and the two-channel model. On one hand, the two-channel model corrects the effect of the out-of-plane phonon at a low computational cost. On the other hand, the introduction of the induced *xy* potential in the electron-phonon calculation indirectly corrects the effect of the band diagram. The $\lambda_{11}^{111}$ is controlled by the band diagram, which contains the information about the effective mass. The numerator and denominator in $R_g$ are obtained from the same band diagram, so that the error due to the effective mass in these three non-heavy fermion superconductors can almost be cancelled.

It is still an open question which DFT functional is the best for iron-based superconductors. From an empirical point of view, the GW or screened hybrid functional is likely suitable for the unconventional bismuthate and the transition-metal superconductors [56]. The modelling of the Hubbard potential in the GGA+U approach provides a good agreement with the experimental results of BaFe$_2$As$_2$ and LaFeAsO [57]. Since the electron-electron interaction in the iron-based superconductors is complicated, the use of the highly correlated DFT functional should be reasonable. However, the $T_c$ calculated with the screened hydrid functional HSE06 convinces us to use a different approach. We calculate the $T_c$ of these three materials by the HSE06 functional, which is a class of approximations to the exchange–correlation energy functional in density functional theory, which includes a part of the exact exchange item from the Hartree–Fock theory with the rest of the exchange–correlation energy from other sources [57]. However, the exchange-correlation energy considered by the screened hydrid functional HSE06 does not suit the NaFeAs, LiFeAs and FeSe materials whose calculated $T_c$ values become less than 0.1K. The more advanced approaches, such as GW or DMFT, can simulate most of the electronic properties of bulk FeSe closer to the experimental values but the major drawback is that the calculation of the electron–phonon coupling with these methods is based on a simplified deformation potential approximation, since electron–phonon coupling matrix elements are difficult to obtain [40].

The induced *xy* potential was rarely reported at GGA level. If the channels where the out-of-plane phonon cannot be hidden are considered separately, the GGA functional is already good enough to generate the induced *xy*-potential. If the lattice Fe moves orthogonally away from the *xy* plane in the iron-based superconductors, the electric charges in the *xy* plane are disturbed. Since the electronegativity of the tetrahedral atom (Se or As) is stronger, the electron will populate the FeSe or FeAs bonds more [21]. For example, when the Fe moves along the +*z* axis, the local electron density in the *xy*-plane changes. The induced charges have two possible paths, i.e. the electrons are shifted either above or below the *xy* plane to the FeSe (or FeAs) bond [21]. However, the upward displacement of the Fe atom, which emits the electric field, confines the electrons more covalently in the upper tetrahedral region. The more covalently bonded FeSe (or FeAs) interaction allows electrons to move out of the FeSe or (FeAs) bond below the plane [21]. A charge fluctuation is created and generates the induced *xy* potential. Since the out-of-plane phonon is simulated by the two-channel model, the occurrence of the induced *xy* potential at GGA level means that the two-channel model has already taken the AF into account.

The McMillian formula takes into account the distribution of electrons in the form of a hyperbolic tangent (*tanh*) function across the Fermi level [41]. At finite temperature, the Fermi-Dirac statistics fits the shape of the hyperbolic tangent function with the mean occupation number $f(E_F) = 0.5$. For example, elemental aluminum holds the superconducting transition temperature at 1.2K, where the offset $f(E_F - E_{Debye}) - f(E_F + E_{Debye})$ is 0.0056. In addition, the offset $f(E_F - E_{Debye}) - f(E_F + E_{Debye})$ of elemental tin is 0.0028 at ~3K. The McMillian formula provides the theoretical $T_c$ of aluminum and tin correctly with the tiny offsets of 0.0056 and 0.0028, respectively. The relevant electrons in the studied superconductors may be located in the energy range between $E_F - E_{Debye}$ and $E_F + E_{Debye}$, but their offsets $f(E_F - E_{Debye}) - f(E_F + E_{Debye})$ at low temperatures are as small as ~0.005. If $f(E_F - E_{Debye}) - f(E_F + E_{Debye})$ in the iron-based superconductors are comparable to BCS superconductors, the numerical error due to the fitting of the relevant electrons indicated by the energy range we extracted from ARPES data as input in the McMillian formula and the Eliashberg function may not be obvious. If the $R_g$ factor is introduced in a narrow energy range below the Fermi level, it fits even better to the *tanh* function. Furthermore, the AF Ising model shows that the energy of the spin fluctuations is smaller than the Debye energy and hence the maximum integral in the McMillian derivation [41] cannot exceed the Debye temperature. Finally, none of the amplified electron-phonon couplings exceeds the limit of the straight-line fit for determining the empirical parameters [41]. Therefore, the McMillian formula becomes applicable in these three iron-based superconductors.

After we consider all electrons taking part in iron-based superconductivity between $E_F$ and $E_F - E_D$, the $T_c$ calculations of the above samples becomes accurate. We thus suggest that, given the relatively high transition temperatures of Fe-based superconductors at which a considerable amount of high energy phonons are excited, it is absolutely required to consider the entire energy range of electrons that can scatter up to the Fermi energy through these phonons, in contrast to the traditional low-$T_c$ approaches, where the electronic density of states at the Fermi level can be used as an approximation. For a proposed theory of iron-based superconductors to be deemed incorrect, an unified theory of iron-based superconductors would need to have already existed. However, what is the unified theory of iron-based superconductor? It is still an open question. Despite our algorithm can produce the theoretical $T_c$ of these four samples at reasonable values, this article only proposes possible pairing mechanisms for the studied samples instead of announcing a theory of iron-based superconductors. But our work give hope to scientist that successful Tc calculations in iron-based superconductors may be achievable. Further theoretical work is still massively required to search for an unified theory of iron-based superconductors that can estimate the theoretical $T_c$ of all iron-based superconductors precisely.

## 5. Conclusion

After revising the superconducting electron-concentration in the McMillan $T_c$ formula, we could show that when the conduction electrons interact with local Fe moments in Fe-based superconductors, the coexistence of superconductivity with local fluctuating antiferromagnetism together with the abnormal lattice vibration, which can lead to an enormous increase in the electron phonon coupling is sufficient to predict the high $T_c$ values. Our ab-initio approach can generate theoretical $T_c$ values of NaFeAs, LiFeAs and FeSe close to the experimental values. When the model applied to monolayered FeSe on a $SrTiO_3$ substrate, we find that the interfacial phonons are of major importance to explain the high-temperature superconductivity.

**Acknowledgements**
We thank Prof. Steven G. Louie in UC Berkeley Physics for his valuable suggestions

**Data Availability Statement:**
Data are sharable under a reasonable request. The authors are usually supportive for reproducing the results if further assistance is needed (Please send your technical requests to roywch654321@gmail.com)